\documentclass[twocolumn,pre,superscriptaddress]{revtex4}

\usepackage{dcolumn}
\usepackage{amsmath}
\usepackage{amssymb}
\usepackage{multirow}

\usepackage{graphicx}

\begin{document}

\newcommand{\half}{\mbox{$\frac12$}}
\renewcommand{\d}{{\rm d}}
\renewcommand{\i}{{\rm i}}
\renewcommand{\O}{{\rm O}}
\newcommand{\e}{{\rm e}}
\newcommand{\set}[1]{\lbrace#1\rbrace}
\newcommand{\av}[1]{\langle#1\rangle}
\newcommand{\eref}[1]{(\ref{#1})}
\newcommand{\etal}{{\it{}et~al.}}
\newcommand{\cosec}{\mathop{\mathrm{cosec}}\nolimits}
\newcommand{\cN}{\mathcal{N}}
\newcommand{\mt}{\widetilde{m}}
\newcommand{\kt}{\tilde{k}}
\newcommand{\kbar}{\bar{k}}
\newcommand{\mbar}{\overline{m}}
\newcommand{\zbar}{\bar{z}}
\newcommand{\Knn}{\bar{K}^\mathrm{nn}}
\newcommand{\knn}{\bar{k}^\mathrm{nn}}
\newcommand{\kbarout}{\bar{k}^\mathrm{out}}
\newcommand{\kbarin}{\bar{k}^\mathrm{in}}
\newcommand{\prodvw}{\prod_{(v,w)}}
\newcommand{\sumvw}{\sum_{(v,w)}}

\newlength{\figurewidth}
\ifdim\columnwidth<10.5cm
  \setlength{\figurewidth}{0.95\columnwidth}
\else
  \setlength{\figurewidth}{10cm}
\fi
\setlength{\parskip}{0pt}
\setlength{\tabcolsep}{6pt}
\setlength{\arraycolsep}{2pt}

\title{The origin of degree correlations in the Internet and other networks}
\author{Juyong Park}
\affiliation{Department of Physics,
University of Michigan, Ann Arbor, MI 48109}
\affiliation{Michigan Center for Theoretical Physics,
University of Michigan, Ann Arbor, MI 48109}
\author{M. E. J. Newman}
\affiliation{Department of Physics,
University of Michigan, Ann Arbor, MI 48109}
\affiliation{Center for the Study of Complex Systems,
University of Michigan, Ann Arbor, MI 48109}
\begin{abstract}
It has been argued that the observed anticorrelation between the degrees of
adjacent vertices in the network representation of the Internet has its
origin in the restriction that no two vertices have more than one edge
connecting them.  Here we introduce a formalism for modeling ensembles of
graphs with single edges only and derive values for the exponents and
correlation coefficients characterizing them.  Our results confirm that the
conjectured mechanism does indeed give rise to correlations of the kind
seen in the Internet, although only a part of the measured correlation can
be accounted for in this way.
\end{abstract}
\maketitle

\section{Introduction}
The statistical properties of networks have been the topic of considerable
attention in the physics literature in recent
years~\cite{Strogatz01,AB02,DM02,Newman03d}.  Motivated by the availability
of large-scale structural data for networks including the Internet, the
World Wide Web, and social and biological networks of various kinds,
researchers have created a wide selection of models of networks and
processes taking place on networks.  One topic of particular current
interest is the issue of degree correlations in networks.  A network or
graph is in general composed of some set of nodes or ``vertices'' joined
together by lines or ``edges,'' and the degree of a vertex is defined to be
the number of edges connected to the vertex.  It has been found that for
many real-world networks the degrees of the vertices at either end of an
edge are not independent, but are correlated with one another, either
positively or negatively~\cite{PVV01,MSZ03,Newman02f}.  A network in which
the degrees of adjacent vertices are positively correlated is said to show
assortative mixing by degree, whereas a network in which they are
negatively correlated is said to show disassortative mixing.  A striking
pattern that emerges when networks of different types are compared is that
most social networks appear to be assortatively mixed, whereas most
technological and biological networks appear to be
disassortative~\cite{Newman02f,Newman03c}.

Of particular interest to us in this paper is the Internet.  At the time of
writing, the Internet forms a network of about $11\,000$ vertices and
$32\,000$ edges, and, as first pointed out by
Pastor-Satorras~\etal~\cite{PVV01}, the degrees of adjacent vertices have
significant anticorrelation.  This is demonstrated by calculating the mean
degree $\knn_v$ of the neighbors of a vertex~$v$ in the network as a
function of the degree~$k_v$ of that vertex.  The resulting function is
found to fall off with increasing~$k_v$ roughly as a power-law $k_v^{-\nu}$
with exponent $\nu\simeq0.5$, so that the higher the degree~$k_v$ of the
one vertex, the lower the mean degree of its neighbors.

In a recent paper, Maslov~\etal~\cite{MSZ03} have proposed a possible
explanation for this result.  Rather than supposing the anticorrelation of
vertex degrees to be the result of some specific social or engineering
constraints on the construction of data networks, they suggest instead a
topological explanation.  Using computer simulations, they show for a
network of the size and degree sequence of the Internet that the
requirement that there is at most one edge between any pair of vertex
induces degree anticorrelations very similar to those observed.  And indeed
there are no double edges in the Internet, a statistically unlikely
occurrence were we given complete freedom about how vertices were
connected.

The physical intuition behind the suggestion of Maslov~\etal\ is that the
restriction to single edges causes high-degree vertices to have fewer
connections between them than they would if edges were assigned purely at
random, and hence there must be more connections between
high-degree/low-degree vertex pairs instead.  A similar explanation could
apply in the case of other types of networks as well, such as directed
networks.  The World Wide Web and foodwebs are two examples of directed
networks that appear to be disassortative and usually have no double
edges~\footnote{The Web can have more than one link from one page to
another, but in most studies of the topology of the Web graph duplicate
links have been neglected.}.

In this paper we study the mechanism proposed by Maslov~\etal\
analytically, and demonstrate that it does indeed produce disassortative
mixing by degree of precisely the type observed by
Pastor-Satorras~\etal~\cite{PVV01}.  The particular model chosen by
Maslov~\etal\ to test their idea turns out to be difficult to treat
analytically.  They studied the ensemble of all graphs with a particular
degree sequence and at most one edge between any vertex pair, in which each
allowed graph appears with equal probability.  Calculating the correlations
in this ensemble requires us to enumerate binary matrices with given row
and column sums.  No closed-form solution for such an enumeration is known
at present despite decades of study by mathematicians~\cite{BC78,WZ98}.  In
this paper, therefore, we take a different approach, borrowing a trick from
statistical mechanics.  We study an expanded ``grand canonical'' ensemble
of graphs in which the number of edges is allowed to vary under the action
of a chemical potential.  As network size becomes large, the number of
edges becomes narrowly peaked and the predictions of the model become
similar to those of the model of Maslov~\etal, while the calculations are
far easier.  (A grand canonical ensemble of graphs has also been studied
recently by Dorogovtsev~\etal~\cite{DMS03a}, although using a different
formalism and to a different purpose.)

For networks with power-law degree distributions, we will show that indeed
$\knn_v$ falls off as a power of~$k_v$ and derive the value
of the exponent~$\nu$.  We also calculate the value of the degree
correlation coefficient for adjacent vertices, which measures the amount of
disassortative mixing in the network.  We show that the mechanism of
Maslov~\etal\ can account for some, but not all, of the disassortativity
seen in the Internet, suggesting that there are also other mechanisms
contributing to the observed degree correlations.

\section{Definitions}
The classic model in the study of graphs with arbitrary degree sequences is
the so-called configuration
model~\cite{BC78,Bollobas80,Luczak92,MR95,MR98,NSW01}, in which one
specifies the degree $k_v$ of each vertex~$v=1\ldots n$ in a network, which
also fixes the total number of edges to be $m=\half\sum_v k_v$.  Subject to
the given degree sequence, the vertices are randomly wired to one another.
The combinatorics of this model are however awkward and so Chung and
Lu~\cite{CL02a} have recently proposed an alternative model that is in many
ways more convenient.  (Models similar to that of Chung and Lu have also
been introduced independently by Dorogovtsev~\etal~\cite{DMS03a} and
Caldarelli~\etal~\cite{CCDM02}.)  As we will show, by making use of an
extension of their model we can make tractable the problem of counting
graphs with single edges only.  The model of Chung and Lu deals with
undirected networks, and we consider that case first. A fairly
straightforward generalization to directed networks will be dealt with
briefly.

\subsection{The network model of Chung and Lu}
In the model of Chung and Lu~\cite{CL02a} one specifies the \emph{desired}
degrees $\kt_v$ of vertices~$v$ and then places edges between vertex pairs
$(v,w)$ with probability
\begin{equation}
f_{vw} = {\kt_v\kt_w\over 2\mt},
\label{defspij}
\end{equation}
where $\mt=\half\sum_v\kt_v$ is the desired number of edges in the graph.
The expected degree of vertex~$v$ is then
\begin{equation}
\kbar_v = \sum_{w} f_{vw} = \frac{\kt_v}{2\mt} \sum_{w}\kt_w = \kt_v.
\label{expectedkv}
\end{equation}
Thus the expected degree of each vertex is equal to its desired degree and
the expected degree distribution is asymptotically equal to the
distribution of the desired degree sequence, although any individual vertex
may have a degree that differs from its desired value.  (Throughout this
paper, we denote desired values of quantities by a tilde (e.g.,~$\kt$),
expected values or ensemble means by a bar (e.g.,~$\kbar$), and actual
values in a particular graph by undecorated characters (e.g.,~$k$).)

However, this approach is not entirely satisfactory.  For some degree
distributions the probability $f_{vw}$ can exceed one.  Physically, this
means that there can be more than one edge between a pair of vertices,
precisely the situation that we will want to exclude in our calculations.
Chung and Lu circumvent this problem by specifying an additional constraint
on the distribution of desired degrees, $\kt_{v} \leq \sqrt{2\mt}$ for
all~$v$.  While this condition ensures that $f_{vw}\le1$, it is strongly
violated by networks like the Internet that have power-law degree
distributions.  Here, therefore, we adopt an alternative strategy, and
adapt the model of Chung and Lu to incorporate an explicit condition that
there is only one edge between every vertex pair.  As we will see, this
leads to some interesting new physics, and in particular to an explanation
of the origin of disassortativity.

\subsection{Ensemble of networks with single edges}
We consider explicitly an ensemble of networks in which there is only a
single edge between any pair of vertices.  There will be an edge between
the pair $(v,w)$ with probability $f_{vw}$ or not with probability
$1-f_{vw}$.  Then the probability of occurrence of a particular graph~$G$
can be written
\begin{equation}
\Gamma(G) = \prodvw (1-f_{vw}) \prod_{\textrm{edges}}
            {f_{vw}\over1-f_{vw}}
\end{equation}
where the first product is over all unique vertex pairs $(v,w)$ and the
second is over only those pairs between which there is an edge.  For
convenience, we will write $P_{vw}=f_{vw}/(1-f_{vw})$, $\Gamma_0=\prodvw
(1-f_{vw})$, and define $\delta_{vw}$ to be 1 if there is an edge between
$v$ and $w$ and zero otherwise.  Then
\begin{equation}
\Gamma(G) = \Gamma_0 \prodvw P_{vw}^{\delta_{vw}}.
\end{equation}

To progress, we need to choose a form for $P_{vw}$, or equivalently for
$f_{vw}$.  We here make the particular choice of the factorizable form
\begin{equation}
P_{vw} = \beta \lambda_v \lambda_w,
\label{pvwchoice}
\end{equation}
where $\beta$ is a free parameter that will control the total number of
edges in the graph, and the fugacity~$\lambda_v$ is a real number assigned
to each vertex~$v$ that will control the expected degree of that vertex.
(Note that this choice is not the same as that of Chung and Lu,
Eq.~\eref{defspij}, although in the ``classical limit'' of graphs with few
double edges it becomes the same.  See below.)

The justification for our choice of $P_{vw}$ is that we would like all
graphs with a given degree sequence to appear in our ensemble with equal
probability.  This is the same criterion applied by
Maslov~\etal~\cite{MSZ03} in their simulations, and allows us to compare
our results with theirs.  To see that the criterion is satisfied in this
case, we need only observe that
\begin{equation}
\Gamma(G) = \Gamma_0 \beta^m \prod_v \lambda_v^{k_v},
\label{defsgammag}
\end{equation}
where $k_v$ is the actual degree of~$v$ in the particular graph~$G$ and
$m=\half\sum_v k_v$ is the actual number of edges.  Clearly for given
$\beta$ and $\set{\lambda_v}$ this expression is a function only of the
degree sequence~$\set{k_v}$.

\section{Predictions and results}
We now define a grand partition function
\begin{equation}
Z = \sum_G \Gamma(G)
  = \Gamma_0 \! \sum_{\set{\delta_{vw}}} \prodvw P_{vw}^{\delta_{vw}}.
\end{equation}
Interchanging the order of sum and product this gives
\begin{eqnarray}
Z &=& \prodvw \sum_{\delta_{vw}} P_{vw}^{\delta_{vw}}
   =  \prodvw (1 + P_{vw})\nonumber\\
  &=& \prodvw (1 + \beta\lambda_v\lambda_w),
\label{defpart}
\end{eqnarray}
where we have dropped the factor of~$\Gamma_0$.  (As is typically the case
with partition functions, leading factors of this type cancel out of all
observable quantities in the theory.)

From Eq.~\eref{defsgammag} we can now see that the expected degree
$\kbar_v$ of vertex~$v$ will be given by
\begin{equation}
\bar{k}_v = \frac{\lambda_v}{Z} {\partial Z\over\partial\lambda_v}
          = -\lambda_v \frac{\partial F}{\partial \lambda_v},
\label{ourkv}
\end{equation}
where $F$ is the free energy
\begin{equation}
F = -\log Z
  = -\sumvw \log(1+\beta \lambda_v \lambda_w).
\label{defsf}
\end{equation}
Combining Eqs.~\eref{ourkv} and~\eref{defsf}, we then get
\begin{equation}
\bar{k}_v = \sum_w
            \frac{\beta\lambda_v\lambda_w}{1+\beta\lambda_v\lambda_w}.
\label{reskv}
\end{equation}

The expected number of edges $\mbar$ is the ensemble mean of the
exponent of~$\beta$ in the partition function, which is given by
\begin{equation}
\mbar = -\beta {\partial F\over\partial\beta}
   = \sumvw \frac{\beta\lambda_v\lambda_w}{1+\beta\lambda_v\lambda_w}.
\label{resm}
\end{equation}
The mean degree of the entire system $\zbar$ is simply $2\mbar/n$,
where $n$ is the total number of vertices.

There are clear parallels between these results and the familiar Fermi
ensemble of elementary statistical mechanics.  The quantity $f_{vw}$
introduced earlier, which we can now write in the form
\begin{equation}
f_{vw} = {\beta\lambda_v\lambda_w\over 1+\beta\lambda_v\lambda_w},
\end{equation}
lies strictly in the range from 0 to~1, and represents the probability that
an edge lies between a particular pair of vertices.  This is the equivalent
of the Fermi function of statistical mechanics.

The mean sum of the degrees of the neighbors of a vertex~$v$, which we
denote $\Knn_v$, is given by
\begin{equation}
\Knn_v = \sum_w f_{vw} \bar{k}_w
       = \sum_w \frac{\beta\lambda_v\lambda_w}{1+\beta\lambda_v\lambda_w}
         \kbar_w,
\label{resknn}
\end{equation}
with $\kbar_w$ given by Eq.~\eref{reskv}, and the mean degree of a neighbor
of~$v$ is equal to $\knn_v=\Knn_v/\kbar_v$.  We will also want to calculate
the correlation coefficient of the degree of vertices at either end of an
edge~\cite{Newman02f}, whose value is given by
\begin{equation}
r =
    \frac{\sum_{v}\kbar_v \Knn_v - (2\bar{m})^{-1}
      \bigl[\sum_{v}\kbar_v^{2} \bigr]^2}{\sum_{v}
      {\kbar_v}^3-(2\bar{m})^{-1} \bigl[\sum_{v}\kt_v^{2} \bigr]^2}.
\end{equation}

Although in this paper we will be dealing primarily with undirected
networks, generalization of the theory to directed networks is
straightforward.  If $f_{vw}$ denotes the probability of existence of a
directed edge from $v$ to~$w$ and $P_{vw}$ is defined as before, then the
expected out-degree (number of outgoing edges) of a vertex~$v$ will be
\begin{equation}
\kbarout_v = \sum_w f_{vw} = \sum_w \frac{P_{vw}}{1+P_{vw}},
\end{equation}
the expected in-degree (number of incoming edges) will be
\begin{equation}
\kbarin_v = \sum_w f_{wv} = \sum_w \frac{P_{wv}}{1+P_{wv}},
\end{equation}
and the obvious generalizations of Eqs.~\eref{resm} and~\eref{resknn}
apply.

\subsection{Example: power-law degree distribution}
We are here particularly interested in the case of the Internet, which,
like a number of other networks, has a degree distribution that
approximately follows a power law
\begin{equation}
p_k \propto k^{-\tau},
\end{equation}
with $\tau\sim2.2\pm0.3$~\cite{FFF99,Chen02}.  The long tail of the power
law means that the highest degree vertex pairs in the network would be
quite likely to have more than one edge running between them were edges
assigned at random, and the behavior of the network changes substantially
when these multiple edges are disallowed.  This is the origin of the
effects observed by Maslov~\etal~\cite{MSZ03} in their simulations.

As we now show, the power-law degree distribution can be reproduced in our
model by choosing the fugacity~$\lambda$ also to have a power-law
distribution with the same exponent~$\tau$, so that the number of vertices
with fugacity between $\lambda$ and $\lambda+\d\lambda$ is
$p(\lambda)\>\d\lambda$, where
\begin{equation}
p(\lambda) = \biggl\lbrace\begin{array}{ll}
         C \lambda^{-\tau} & \qquad\mbox{for $\lambda\ge\lambda_0$} \\
         0                 & \qquad\mbox{for $\lambda<\lambda_0$.}
             \end{array}
\label{plambda}
\end{equation}
The lower cutoff makes the distribution normalizable, and $C$ is a
normalizing constant given by
\begin{equation}
C^{-1} = \int_{\lambda_0}^\infty \lambda^{-\tau} \>\d\lambda
       = {\lambda_0^{-\tau+1}\over\tau-1}.
\end{equation}
(Bear in mind that $\lambda$ is not restricted, as the degree is, to
integer values.)

Let us consider the case $\tau=\frac52$, for which the expressions for the
quantities of interest take particularly simple forms.  For this choice,
the expected degree $\kbar(\lambda)$ of a vertex with fugacity $\lambda$ is
\begin{eqnarray}
\hspace{-2em}
\kbar(\lambda) &=& n \int_{\lambda_0}^{\infty} \frac{\beta \lambda
	 \lambda'}{1+\beta \lambda \lambda'}\,p(\lambda')\>\d\lambda'
	 \nonumber \\ &=& 3n \bigl[ \beta\lambda_0\lambda - (\beta
	 \lambda_0\lambda )^{3/2} \arctan \bigl(
	 (\beta\lambda_0\lambda)^{-1/2} \bigr) \bigr],
\label{kmean}
\end{eqnarray}
and the mean degree $\zbar$ of the system is
\begin{eqnarray}
\zbar &=& \frac{2\mbar}{n}
= 2 \int_{\lambda_0}^{\infty}
  \kbar(\lambda)\,p(\lambda)\>\d\lambda \nonumber\\
&=& 9n\beta{\lambda_0}^2 \biggl[ 1 - \mbox{$\frac14$}\Phi\biggl(
          -\frac{1}{{\beta \lambda_0}^2},2,\half \biggr) \biggr],
\label{zmean}
\end{eqnarray}
where $\Phi(x,a,b)$ denotes the analytic continuation of the Lerch
transcendent. 

The parameter $\beta$ is to some extent redundant in these expressions,
since we are free to choose $\lambda$ as we wish, but it proves convenient
nonetheless.  If we choose $\beta=(2\mt)^{-1}$, where $\mt$ is the desired
number of edges as before, then for graphs in which there are few double
edges we have $f_{vw} \simeq (\lambda_v\lambda_w)/(2\mt)$, giving
$\bar{k}_v = \sum_{w}f_{vw} \simeq \sum_w (\lambda_v\lambda_w)/(2\mt)
\simeq \lambda_v$, so that the fugacity is simply equal to the desired
degree of a vertex, as in the model of Chung and Lu~\cite{CL02a}.

The regime in which there are few double edges can be thought of as the
``classical limit'' of our Fermi ensemble, and corresponds to the case
where the first terms in Eqs.~\eref{kmean} and~\eref{zmean} dominate.  As
$\lambda$ becomes large, however, encouraging vertices to have high degree,
we enter the quantum regime, where it becomes harder and harder for
vertices to find others to connect to.  This is reflected in
Eq.~\eref{kmean} also.  Expanding the inverse tangent as $\arctan
x=x-\frac13 x^3+\frac15 x^5-\O(x^7)$, we find that the leading term cancels
and
\begin{equation}
\kbar(\lambda) = n - {3n\over5\beta\lambda_0\lambda}
                 + n\O\bigl((\beta\lambda_0\lambda)^{-2}\bigr).
\end{equation}
Thus, as $\lambda\to\infty$ the degree tends to~$n$, as we would expect,
since this is the largest degree a vertex can have on a network with no
double edges.


The mean sum $\Knn(\lambda)$ of the degrees of the neighbors of a vertex
with fugacity~$\lambda$ is
\begin{eqnarray}
\Knn(\lambda)
 &=& n \int_{\lambda_0}^{\infty}
       \frac{\beta\lambda\lambda'}{1+\beta\lambda\lambda'}
       \kbar(\lambda')\,p(\lambda') \>\d\lambda' \nonumber\\
 &=& 9 n^2 (\beta\lambda_0\lambda)^{3/2}
       \biggl[ {\lambda_0\over\lambda} \arctan \bigl(
       (\beta\lambda_0\lambda)^{-1/2} \bigr) \nonumber\\
 & & {} - \frac{\pi}{4} \biggl({\lambda_0\over\lambda}\biggr)^{3/2}
     \bigl[ 2\log\bigl(1+(\lambda/\lambda_0)^{1/2}\bigr) \nonumber \\
 & & \hspace{2em}{}-  \log (1+\beta\lambda_0\lambda)+O(\lambda_0 \beta^{\frac{1}{2}})
     \bigr]\biggr],
\label{Knn}
\end{eqnarray}
and from this we can calculate~$\knn$~\footnote{To be precise, we note that
our formulas give $\knn$ as a function of the expected degree of a vertex,
rather than its actual degree.  However in the large-degree tail, where the
power-law behavior of interest is observed, we expect that actual degree
will be narrowly peaked around the expected value.}.

These results can be extended to other values of $\tau$ also, although the
formulas are not as elegant as for the case $\tau=\frac52$.  For example,
for general $\tau>1$ the equivalent of Eq.~\eref{kmean} is
\begin{equation}
\kbar(\lambda) = n\,
  {_2}F_1\biggl(1,-1+\tau;\tau;-\frac{1}{\beta\lambda_0\lambda} \biggr),
\end{equation}
where ${_2}F_1$ is a hypergeometric function.  This form is used in some of
the calculations in the next section.

\begin{figure}
\resizebox{\figurewidth}{!}{\includegraphics{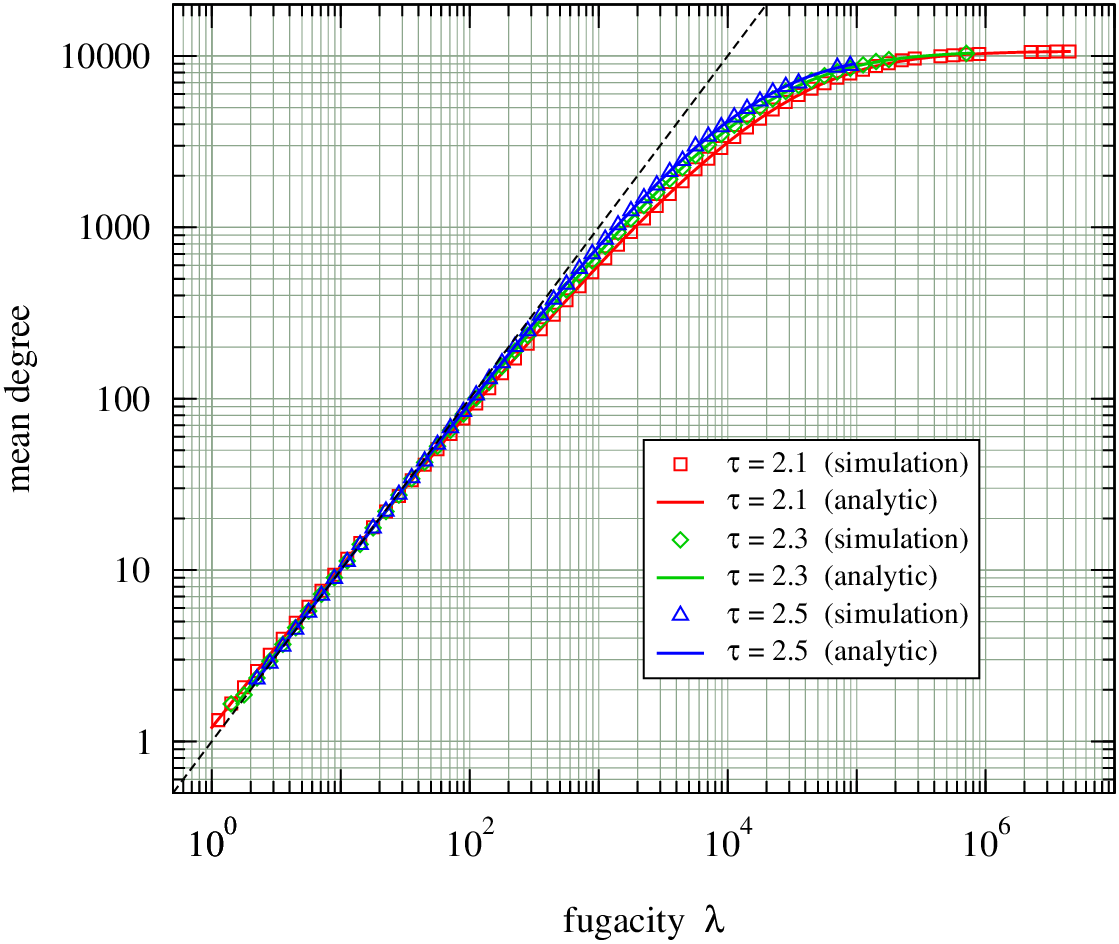}}
\caption{The ensemble mean $\kbar$ of the degree of a vertex in our model
as a function of the fugacity~$\lambda$ of the vertex.  The numerical
results are averaged over 1000 repetitions of the simulation.  The dotted
line indicates the form $\kbar=\lambda$, which the curve is expected to
approximate for small~$\lambda$.}
\label{kvsl}
\end{figure}

\subsection{Comparison with the Internet}
\begin{figure}
\resizebox{\figurewidth}{!}{\includegraphics{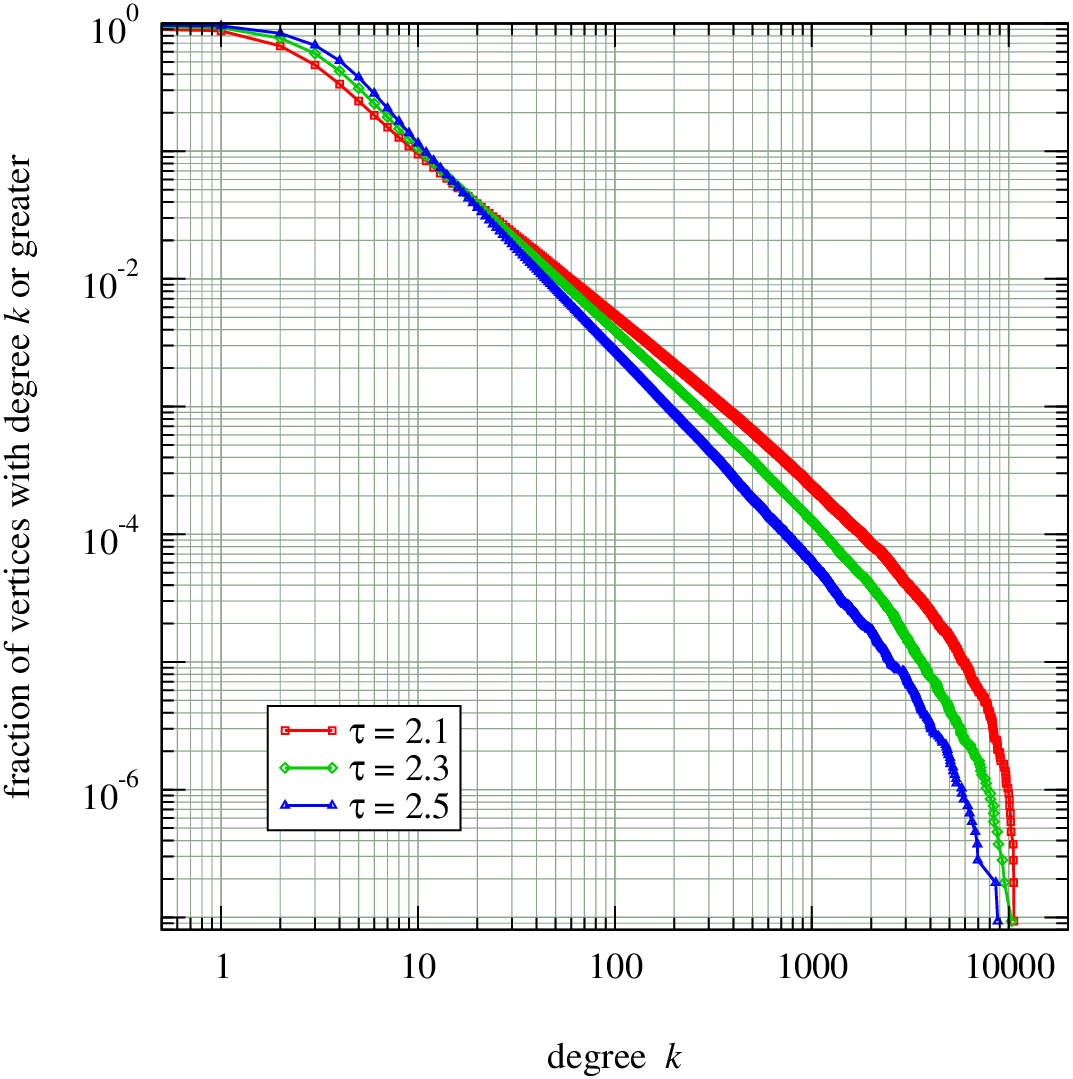}}
\caption{The cumulative distribution function for vertex degree in
simulations of our model.  The general form of the distribution is a power
law for low degree with a cutoff as degree approaches the system size~$n$.}
\label{pkvsk}
\end{figure}

\begin{figure}
\resizebox{\figurewidth}{!}{\includegraphics{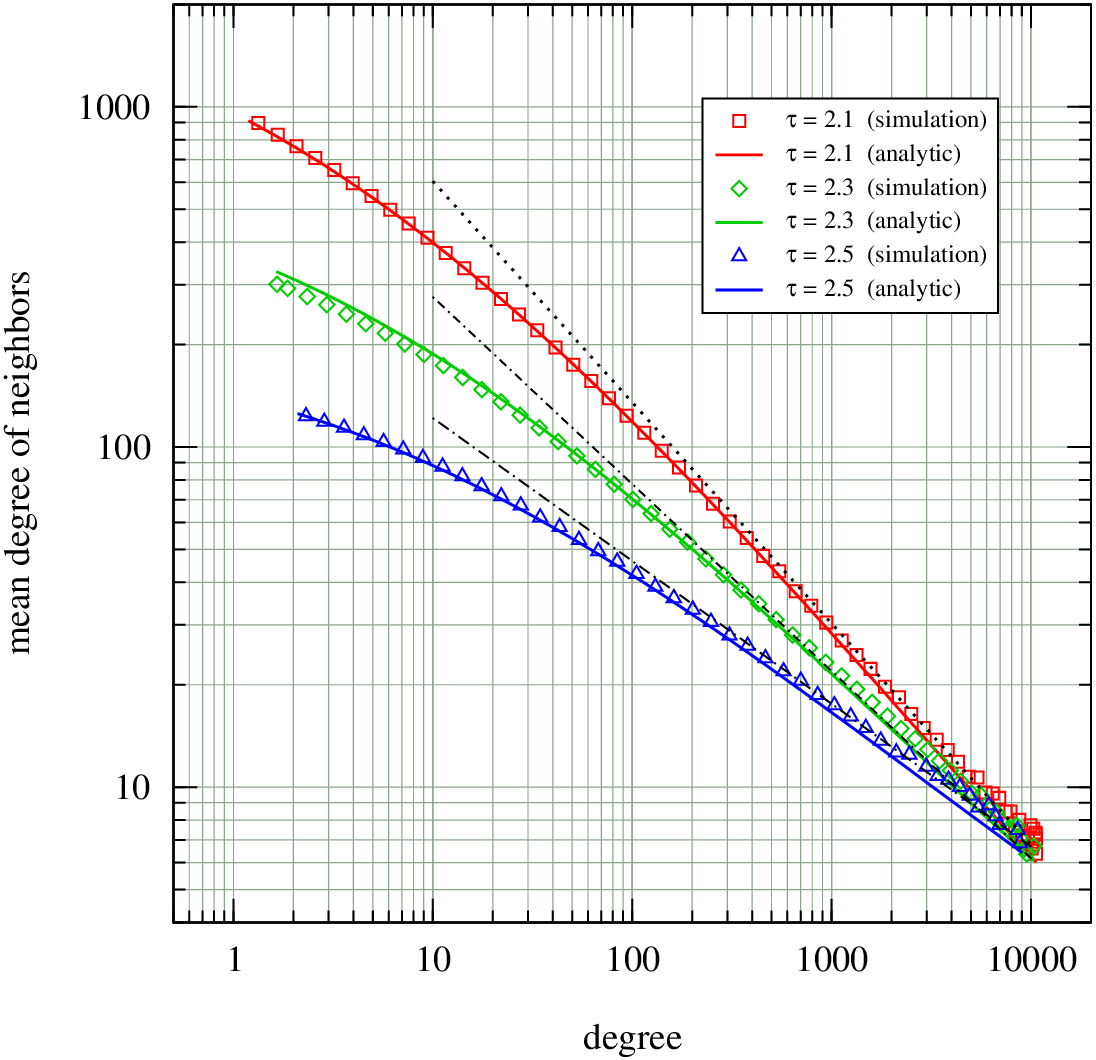}}
\caption{The mean degree $\knn$ of the neighbors of a vertex as a function
of the degree $\kbar$ of that vertex.  The dotted lines show the asymptotic
slopes of the curves.}
\label{knnk}
\end{figure}

We now compare our model quantitatively with the Internet graph.  To do
this, it is important that we make the size~$n$ and number of edges~$\mt$
the same as the real Internet, since our predictions, Eqs.~\eref{kmean}
and~\eref{Knn}, are dependent on these quantities.  For the purposes of
comparison, we use the data of Chen~\etal~\cite{Chen02} from 2001 on the
structure of the Internet at the autonomous system level, for which
$n=10\,697$ and $\mt=31\,992$, which gives a mean degree of
$\zbar=2\mt/n=5.98149$.  For the choice Eq.~\eref{plambda} of fugacity
distribution used here, we can arrange for the network to have the correct
mean degree by an appropriate choice of the lower limit~$\lambda_0$ of the
distribution, and we do this for three values $\tau=2.1$, $2.3$, and $2.5$
of the exponent of the power-law.  We also perform extensive simulations of
the model for the same parameter values to confirm our calculations, and
analytic and numerical results are shown in Figs.~\ref{kvsl},~\ref{pkvsk},
and~\ref{knnk} and in Table~\ref{values}.  As we can see, analytic and
numerical predictions agree closely.

Consider first Fig.~\ref{kvsl}, which is a plot of the mean degree of a
vertex as a function of its fugacity.  As the figure shows, the degree is
closely linear in the fugacity for small~$\lambda$ and flattens off as
degree approaches~$n$, as expected.

The same behavior is evident in Fig.~\ref{pkvsk} also, which shows the
cumulative distribution function of degrees in simulations of the model for
power-law distributed fugacity, Eq.~\eref{plambda}.  The distribution of
degrees also follows a power law (a straight line on the logarithmic axes
used), until degree approaches~$n$, where the distribution is cut off.
This is eminently sensible behavior: given the constraint of single edges
only, presumably the real Internet must deviate from power-law behavior for
large degree, and our model should and does reflect this behavior.

The fundamental result of this paper is shown in Fig.~\ref{knnk}, where we
have plotted the mean degree $\knn$ of the neighbors of a vertex,
calculated from Eq.~\eref{Knn}, against the degree of that vertex.  This is
the comparison used by Pastor-Satorras~\etal~\cite{PVV01} to demonstrate
degree anticorrelation in the Internet.  As the figure shows, there is a
clear decline in the value of $\knn$ as degree increases, just as in the
real Internet, confirming that the single-edge constraint does indeed give
rise to anticorrelations, as conjectured by Maslov~\etal\ \ Furthermore,
the decline appears to be approximately power-law in form
$\knn\sim\bar{k}^{-\nu}$, as found by Pastor-Satorras~\etal\ \ We can
deduce approximate values for the exponent~$\nu$ from our results.  We find
for $\tau=2.1$, $\nu\simeq0.65$, for $\tau=2.3$, $\nu\simeq0.55$, and for
$\tau=2.5$, $\nu\simeq0.42$.  The slopes are shown as the dotted lines in
Fig.~\eref{pkvsk}.  The values for $\nu$ are all close to the value
$\nu\simeq0.5$ observed for the real Internet~\cite{PVV01}.  The power law
is only approximate however---the functional form of Eq.~\eref{Knn} is not
just a simple power law, and we can see from the figure that the slope of
$\knn$ is smaller for smaller $\kbar$.  The same behavior is visible in
both the real Internet data and the simulation results of
Maslov~\etal~\cite{MSZ03}.

Finally, in Table~\ref{values}, we show values for the mean degree~$\zbar$
and degree correlation coefficient~$r$ for our model.  As we see, the
theoretical calculations and numerical results again agree well.  Since the
Internet is disassortative, we expect the degree correlation coefficient to
be negative in the real network, and its value has been measured to be
$r=-0.189$~\cite{Newman02f}.  In the model we also see negative values
of~$r$, whose magnitude depends quite strongly on the value of the
exponent~$\tau$.  A detailed comparison of model and real-world data may
therefore have to wait on more precise measurements of the degree
distribution (about which there is at present some dispute~\cite{Chen02}).
However it is interesting to note that none of the cases in
Table~\ref{values} is as strongly anticorrelated as the real Internet.
Thus our calculations appear to indicate that some of the disassortativity
in the Internet can be accounted for by the mechanism proposed by
Maslov~\etal, but probably not all of it.  The remainder of the
disassortativity is presumably due to engineering or social constraints on
the structure of the network.  One possibility, which has been discussed
elsewhere~\cite{MSZ03,Newman03c}, is that the Internet is divided into
connectivity providers such as phone companies and ISPs, who tend to have
very high degree, since they have lots of customers, and end users of
connectivity, who typically have a degree of only one or two.  Most
connections in the network run from the providers to the end users and are
therefore from high to low degree, giving a social reason for
disassortativity in addition to the purely topological one considered here.

\begin{table}[t]
\begin{tabular}{c|cc|cc}
       & \multicolumn{2}{c|}{mean degree $\zbar$} & \multicolumn{2}{c}{degree correlation $r$} \\
$\tau$ & theory & simulation                      & theory   & simulation \\
\hline
$2.1$ & $5.981$ & $5.982(15)$                     & $-0.0950$ & $-0.0932(17)$ \\
$2.3$ & $5.981$ & $5.972(9)$                      & $-0.0541$ & $-0.0551(18)$ \\
$2.5$ & $5.981$ & $5.986(7)$                      & $-0.0304$ & $-0.0321(14)$
\end{tabular}
\caption{Mean degree and degree correlation coefficient for the networks
generated by our model from both the analytic theory and from computer
simulations.  The simulation results are averaged over 1000 networks each.
Figures in parentheses show statistical errors on the least significant
figures.}
\label{values}
\end{table}

\section{Discussion and conclusions}
In this paper we have studied analytically ensembles of networks where
there is at most one edge between any pair of vertices.  By making use of
an enlarged ensemble in which the number of edges is allowed to vary, in a
manner reminiscent of the Fermi ensemble of traditional statistical
mechanics, we have been able to find closed form expressions for ensemble
averages of a number of quantities of interest.  In particular, we have
confirmed the previous numerical finding~\cite{MSZ03} that graph ensembles
with single edges have negative correlations between the degrees of
adjacent vertices.  This has been proposed as a possible explanation for
the anticorrelation or disassortativity observed in the topology of the
Internet~\cite{PVV01}.  We find that the restriction to single edges only
can account for some but not all of the correlations observed in real
Internet data.

The same mechanism could be responsible for disassortativity in other
networks also.  Many networks, including citation networks, the World Wide
Web, social networks, collaboration networks, metabolic and genetic
regulatory networks, and food webs have, at least in their most common
representations, only single edges between vertex pairs.  Thus it is
reasonable to suppose that these networks would be disassortative also, and
indeed this appears to be the case for most networks that have been
studied~\cite{Newman03c}.  There is one important exception to this rule
however: social networks almost all appear to be significantly
\emph{assortative} in their mixing patterns.  We conjecture, therefore,
that disassortativity by degree is the normal state of affairs for a
network, as a result of the mechanisms described in this paper, with social
networks being assortative probably because of additional social effects
that are absent from other network types; for one reason or another, it
appears that gregarious people prefer to associate with other gregarious
people.  Furthermore, when assessing the level of assortativity in a social
network, one should take into account the natural tendency for networks to
be disassortative, since this tendency implies that to reach a level even
of neutral assortativity would take a moderately strong bias in favor of
positive degree correlation, and reaching a substantially assortative state
would take a very strong such bias.

Finally, we point out that the general analytical technique employed in
this paper, of enlarging an ensemble of graphs to create a ``grand
canonical'' ensemble, may have applications to other problems in the study
of networks also.  It is well known among statistical physicists that using
such an ensemble often makes the analytic treatment of a problem easier,
and the results presented here offer hope that this approach may prove
useful in other settings.

\begin{acknowledgments}
The authors thank Supriya Krishnamurthy for useful discussions and
comments.  This work was supported in part by the National Science
Foundation under grant DMS--0234188 and by the Michigan Center for
Theoretical Physics at the University of Michigan.
\end{acknowledgments}

\end{document}